\documentstyle[12pt,twoside,fleqn,espcrc1,epsfig]{article}

\newcommand{\AmS}{{\protect\the\textfont2
  A\kern-.1667em\lower.5ex\hbox{M}\kern-.125emS}}

\title{Charmonium Production in High Energy Collisions}

\author{Eric Braaten\address{Department of Physics, Ohio State University,
		 Columbus OH, 43210 USA}
        \thanks{This work was supported in part by the United States
        Department of Energy, Division of High Energy Physics, 
        under grant DE-FG02-91-ER40690.}
        }

\begin{document}


\hfill \vbox{\halign{&#\hfil\cr
		     & OHSTPY-HEP-T-96-016 \cr
                     & hep-ph/9608370      \cr
		     & August 1996         \cr
	}}

\bigskip
\noindent
{\large \bf Charmonium Production in High Energy Collisions}\footnote{Invited 
	talk presented at Quark Matter 96 in Heidelberg, May 1996.}

\bigskip
\noindent
Eric Braaten\footnote{This work was supported in part by the United States
        Department of Energy, Division of High Energy Physics, 
        under grant DE-FG02-91-ER40690.}

\medskip
\noindent
Department of Physics, Ohio State University, Columbus OH, 43210 USA
        
\begin{abstract}
\end{abstract}

\section{INTRODUCTION}

One motivation for studying the production of charmonium 
in high energy collisions is that these particles are among 
the simplest probes for heavy-ion collisions.
In recent years, there have been significant developments 
in heavy quarkonium production, both in theory and in experiment.
These developments may have important implications for the use of 
charmonium as a probe in heavy-ion collisions.
A summary of these developments is presented below.
Those interested in more details
are referred to a recent review article by Braaten, Fleming, and  
Yuan \cite{B-F-Y}. 

\section{CHARMONIUM AS A PROBE}

The simplest probes for hard processes in
heavy-ion collisions are leptons and photons.
One can argue that the next simplest are the heavy quarkonium states,
bottomonium and charmonium.
The bottomonium states are too heavy to be produced in abundance
in heavy-ion collisions, but charmonium states are produced 
reasonably copiously.  The states with the cleanest 
experimental signatures are the $J^{PC} = 1^{--}$
states $J/\psi$ and $\psi'$, which decay into $e^+ e^-$ and $\mu^+ \mu^-$.
The $J^{++}$ states $\chi_{cJ}$, $J = 1,2$, also have reasonably good 
experimental signatures through their 
radiative decays into $\psi + \gamma$.

The reason that these charmonium states are such simple probes is that they 
are essentially 2-body systems. The $\psi$, for example, is to a good 
approximation a bound state of a $c \bar c$ pair in a color-singlet state
with angular-momentum quantum numbers $^3 S_1$. The 
color wavefunction is $(R \bar R + B \bar B + G \bar G)/\sqrt{3}$,
the space wavefunction has the form
$\psi({\bf r}) = R(r)/\sqrt{4 \pi}$, where $R(r)$ is the radial wavefunction, 
and the possible spin wavefunctions are 
$\uparrow \uparrow$, $(\uparrow \downarrow + \downarrow \uparrow)/\sqrt{2}$,
and $\downarrow \downarrow$.
This $c \bar c$ state is the dominant component of the wavefunction, 
but there are also higher Fock state components.
There is, for example, a small $c \bar c g$ component.
Since the overall color state of the $c \bar c g$ 
must be color-singlet, the $c \bar c$ pair must
be in a color-octet state.  The 8 independent color-octet states are 
$R \bar G$, $G \bar B$, $B \bar R$, $R \bar B$, $B \bar G$, $G \bar R$, 
$(R \bar R - G \bar G)/\sqrt{2}$, and $(G \bar G - B \bar B)/\sqrt{2}$.
For most observables, the contribution from the $c \bar c g$ component of the 
wavefunction is small.  However, for some observables, the 
contribution from the $c \bar c$ component of the wavefunction
is suppressed and components
in which the $c \bar c$ pair are in a color-octet state can be important.

There are three important features of
charmonium that make it simpler than light hadrons.
For each of these simplifications, there is a theoretical tool
that can be applied to charmonium 
that is not available in the case of light hadrons.
These tools give us a handle on the effects of gluons
in each of the three most important  ranges of wavelength. 
These three ranges of wavelengths are much less than, comparable to, 
and much greater than $r$, where $r$ is the typical separation
of the $c \bar c$ pair in charmonium.
The first simplifying feature of charmonium is that the mass of the 
charm quark ($m_c \approx$  1.5 GeV) is much larger than the momentum 
scale $\Lambda_{QCD}$ associated with nonperturbative effects in QCD.
This allows us to use  perturbation
theory in the running coupling constant $\alpha_s(m_c)$
to calculate the effects of gluons whose wavelengths are 
of order $1/m_c$ or larger.  The second simplifying feature
is that the $c$ and $\bar c$ are nonrelativistic:  the typical value of 
the relative velocity $v$ of the $c$ and $\bar c$
is about ${1 \over 2}$, so $v^2 \approx {1 \over 4}$.  
This allows us to use a relativistic expansion in powers of $v^2$ 
to organize the effects of gluons whose wavelengths 
are comparable to $r$, which is on the order of $1/(m_c v)$.
The third simplifying feature
is that the geometrical size of charmonium states is significantly smaller 
than that of light hadrons:  the typical separation $r$ of the $c$ 
and $\bar c$ is only about 0.2 fm.  This enables us to use the 
multipole expansion in powers of $r/\lambda$ to simplify the interactions 
of gluons whose wavelengths $\lambda$ are much larger than $r$.
By exploiting these three theoretical tools (perturbation theory, 
the relativistic expansion, and the multipole expansion),
we hope to understand charmonium physics in sufficient 
detail to make it a useful probe of heavy-ion collisions.

\section{CHARMONIUM PRODUCTION}

We now turn to the problem of charmonium production, where we will focus 
on the production of the $J/\psi$ in hadron collisions.
We can think of the production of this state as proceeding through 
two steps.  The first step is the creation of a $c \bar c$ pair
with small relative momentum.  
If their relative momentum is much larger 
than $m_c v$, they will be unlikely to bind and will instead fly 
apart and form $D$ mesons. The creation of the $c \bar c$ pair
must involve momenta 
on the order of $m_c$ or larger, because the momenta of the 
incoming particles provide the rest energy of the $c$ and $\bar c$.
The second step in the production of the $\psi$
is the binding of the $c \bar c$ pair.
This step necessarily involves momenta of order 
$m_c v$ or smaller, because gluons whose wavelengths are comparable to the 
size of the bound state play a large role in the binding.

We first consider the creation of the $c \bar c$ pair.  
One way to produce them is through collisions
between  partons from the colliding hadrons.  
The simplest such processes are $g g \to c \bar c$ and 
$q \bar q \to c \bar c$.  Parton collision processes 
such as these always involve virtual
particles that are off their mass-shells by amounts of order $m_c$.
The cross section for creating the $c \bar c$ pair 
can therefore be calculated as a perturbation series in $\alpha_s(m_c)$.
One can also show that the $c \bar c$ pair from such a process
is produced with separation of order 
$1/m_c$ or smaller, which is much smaller than the size 
of a quarkonium state.  Thus, when the  $c \bar c$ pair is 
produced, it is essentially pointlike
on the scale of the charmonium wavefunction.

There are also production mechanisms that produce $c \bar c$ pairs
that are not pointlike.
For example, a gluon from one hadron can undergo a
quantum fluctuation into a virtual $c \bar c$ pair.
This pair can make a transition into a real $c \bar c$ pair
by exchanging soft gluons with the other hadron.
The separation of the resulting $c \bar c$ pair is 
comparable to  the wavelength of the exchanged gluon.
Such processes are particularly important in forward or diffractive
production. Fortunately, one can minimize the complications from 
such processes by going to a kinematic region where they are suppressed,
such as transverse momentum much greater than $\Lambda_{QCD}$.
We will therefore restrict our attention from this point on to 
parton collision processes that produce pointlike $c \bar c$ pairs.

We now consider the binding of the $c \bar c$ pair to form the $\psi$. 
Assuming that the $c \bar c$ pair is produced through parton collisions,
all we need to know is the probability for a pointlike $c \bar c$ pair
to bind to form a $\psi$.  The inclusive differential cross section can 
be expressed in a form in which short-distance and long-distance
contributions have been factored:
\begin{equation}
d \sigma (\psi + X) \;=\; 
\sum_n  d \widehat{\sigma} \left( (c \bar c)_n + X \right) 
	\langle {\cal O}^\psi_n \rangle.
\label{fact}
\end{equation}
The sum over $n$ includes  all possible color  and angular momentum states
of the  $c \bar c$ pair.  The short-distance factor is the cross section 
$d \widehat{\sigma}$ for creating a $c \bar c$ pair in the state $n$, 
which can be calculated as a perturbation expansion in $\alpha_s(m_c)$.
The long-distance factor $\langle {\cal O}^\psi_n \rangle$ gives the 
probability for a pointlike $c \bar c$ pair in the state $n$
to bind to form the $\psi$.
The matrix elements $\langle {\cal O}^\psi_n \rangle$ contain all the 
nonperturbative information that is required to calculate the 
inclusive cross section.

Until the last few years, most of the predictions for charmonium production
in high energy collisions were carried out using the 
{\it color-singlet model} \cite{Schuler}.  This model 
is a very simple prescription for the probability for a pointlike 
$c \bar c$ pair to form a charmonium meson.
The factor $\langle {\cal O}^\psi_n \rangle$ 
is assumed to be nonzero only if  the state $n$ is a color singlet with the 
same angular-momentum quantum numbers as the dominant Fock state of the meson,
which is $^3S_1$ in the case of the $\psi$.
The probability for a pointlike $c \bar c$ pair in
the color-singlet $^3S_1$ state to form the $\psi$ is the same as 
the probability for the $\psi$ to consist of a pointlike $c \bar c$ pair.
It is proportional to $|R(0)|^2$, where $R(0)$ is the radial 
wavefunction at the origin.
The color-singlet model is very predictive.  It gives predictions for
$\psi$ production in all high energy processes in terms of the single 
nonperturbative factor $|R(0)|^2$.
Moreover, the factor $|R(0)|^2$ can be measured experimentally 
by the decay rate for $\psi \to e^+ e^-$.
Thus the predictions of the color-singlet model are absolutely normalized.

We now consider the production of prompt charmonium in $p \bar p$ collisions.
``Prompt'' means that the charmonium is produced by QCD interactions
rather than by weak decays of hadrons containing bottom quarks.
The cross section for prompt $\psi$ production at Fermilab's Tevatron 
has been measured by the CDF collaboration for 
transverse momenta in the range 5 GeV $< \; p_T \; <$ 20 GeV \cite{Sansoni}.
The background from decays of bottom hadrons were removed
using a silicon vertex detector.  The contribution from
the radiative decays $\chi_{cJ} \to \psi \gamma$ was measured 
and subtracted.  The data is shown in Figure 1.  The dotted line gives the 
prediction of the color-singlet model at large $p_T$. (The dashed line 
gives another contribution in the color-singlet model
that becomes important at lower $p_T$.)
The result is rather surprising.  The measured cross section 
was found to be about a factor of 30 larger than predicted 
by the color-singlet model.  

\begin{figure}
	\begin{center}
	\mbox{\epsfig{file=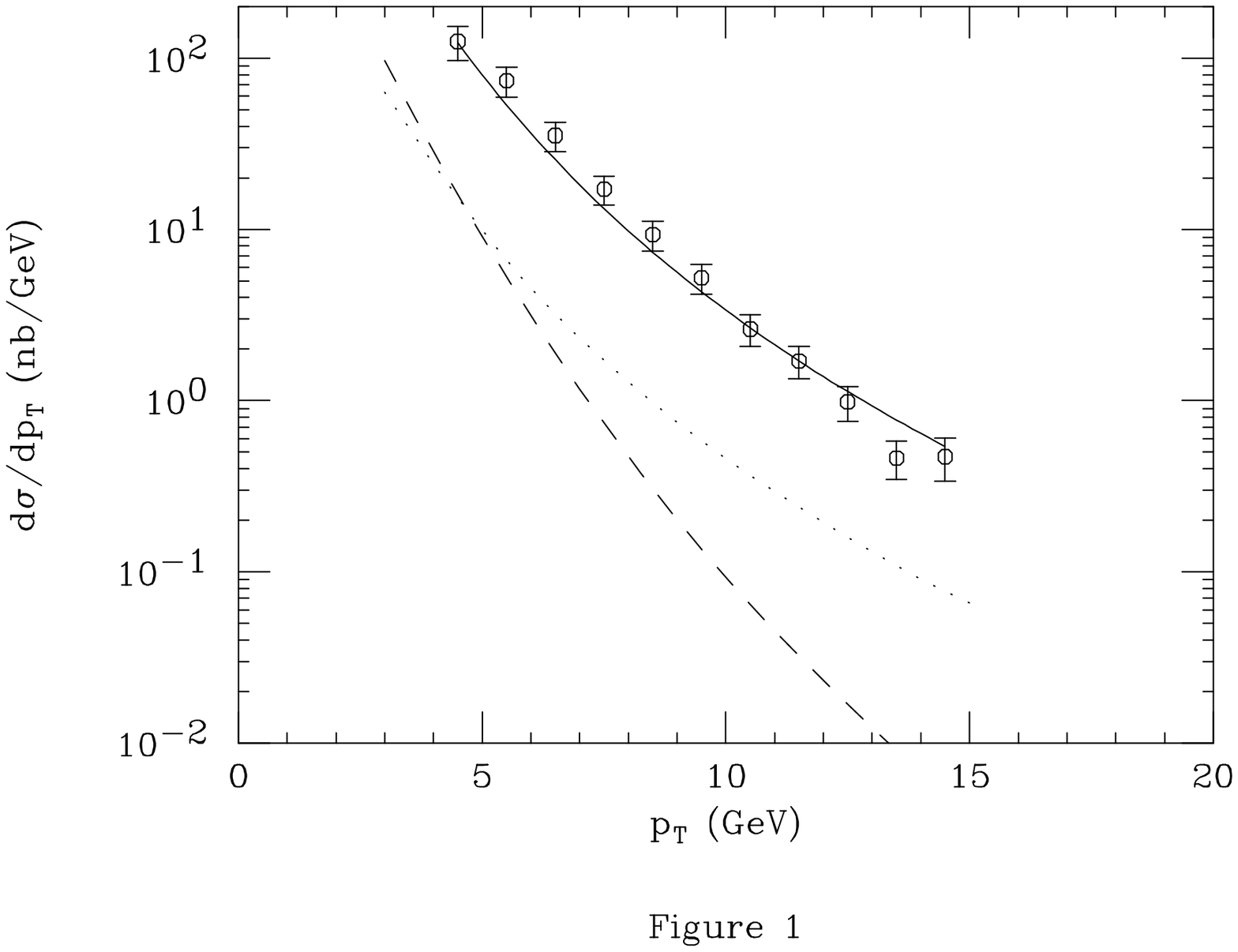, height=10.0cm}}
	\end{center}
\caption{CDF data on the differential cross section
	for prompt $\psi$'s that do not come from $\chi$'s 
	as a function of $p_T$.  The dashed and dotted lines 
	are two contributions predicted by the color-singlet model,
	while the solid line is the prediction of the 
	color-octet mechanism
	with the normalization adjusted to fit the data.}
\end{figure}

This enormous discrepancy between theory and experiment is 
very surprising, because the data extends out to large $p_T$
where the theoretical analysis is particularly clean.
There are factorization theorems of perturbative QCD that 
guarantee that the inclusive differential cross section for the production
of a hadron at large $p_T$ must be dominated by {\it fragmentation}, 
the formation of that hadron in the jet initiated by a parton
with large transverse momentum.  In the case of charmonium,
large transverse momentum means $p_T \gg m_c$ and the most important 
partons are gluons.  The inclusive differential cross section can 
therefore be written in the form
\begin{equation}
d \sigma (\psi(P) + X) \;=\; 
\int_0^1 dz \;   d \widehat{\sigma} \left( g(P/z) + X \right) 
	D_{g \to \psi}(z),
\label{sigfrag}
\end{equation}
where $d \widehat{\sigma}$ is the differential cross section for 
producing a gluon with momentum $P/z$ and $D_{g \to \psi}(z)$
is the {\it gluon fragmentation function}, which 
gives the probability that the jet initiated by the gluon will include a
$\psi$ carrying a fraction $z$ of the gluon momentum. 
All of the dependence on $p_T$ resides in the factor 
$d \widehat{\sigma}$, which  can be calculated using 
perturbation theory  in $\alpha_s(p_T)$.  At leading order, it comes
from the parton process $g g \to g g$, which has a cross section of order
$\alpha_s^2(p_T)$.  All effects from momentum scales of order
$m_c$ or smaller are factored into $D_{g \to \psi}(z)$. 
In particular, the nonperturbative effects associated with the 
binding of the $c \bar c$ pair are all included in 
the fragmentation function.
The color-singlet model gives a definite prediction for the 
fragmentation function in (\ref{sigfrag}). 
The leading contribution comes from the 
parton process $g \to c \bar c g g$.
The resulting fragmentation function is proportional 
to $\alpha_s^3(m_c) |R(0)|^2/m_c^3$.
The dotted curve in Figure 1 is the  prediction obtained using this 
fragmentation function.
Gluon fragmentation gives the dominant contribution in the 
color-singlet model for $p_T$ greater than about 5 GeV.
However, the predicted cross section falls about a factor 
of 30 below the CDF data.

How can we make sense of this enormous discrepancy between theory 
and experiment?
The factorization theorems summarized by (\ref{sigfrag}) are on 
a very firm foundation.  The cross section must certainly
have this form at the largest values 
of $p_T$ that are available. If the prediction fails to agree
with the data, the problem must lie in our assumption about the
fragmentation function $D_{g \to \psi}(z)$.  
The color-singlet model must drastically underestimate the probability 
for a gluon to fragment into charmonium.

\section{NRQCD FACTORIZATION APPROACH}

How can the probability for the gluon to fragment into charmonium be 
so much larger than predicted by the color-singlet model?
There is a simple and natural explanation that
is based on a new approach to charmonium production 
developed recently by Bodwin, Braaten, and Lepage
\cite{B-B-L}.  This approach is called
the NRQCD factorization formalism, because it makes use of an effective 
field theory called nonrelativistic QCD.
The basic idea is to exploit the relativistic expansion for the 
long-distance factors $\langle {\cal O}^\psi_n \rangle$
in the factorization formula (\ref{fact}).  
The factor $\langle {\cal O}^\psi_n \rangle$ gives the probability
for a pointlike $c \bar c$ pair in the color and angular-momentum 
state labelled by $n$ to bind to form a $\psi$.
It can be expressed as a well-defined NRQCD
matrix elements that scales in a definite way with $v$, the 
typical relative velocity of the $c$ and $\bar c$.
The leading matrix element in $v$  
is proportional to $|R(0)|^2$ and scales like $v^3$.
The corresponding term in the factorization formula (\ref{fact}) is 
the cross section of the color-singlet model.
All other terms have matrix elements that are suppressed by powers of $v^2$.
Thus, if the parton cross sections $d \widehat{\sigma}( (c \bar c)_n+X)$
were all comparable, the color-singlet model
would indeed give the dominant term in the cross section.

However, the parton cross sections $d \widehat{\sigma}( (c \bar c)_n+ X)$
can vary widely in magnitude.  They can be suppressed 
not only by powers of $\alpha_s(m_c)$ or $\alpha_s(p_T)$, 
but also by powers of
kinematical factors like $m_c/p_T$.  If the color-singlet model
term is suppressed by such a factor, then other terms may be important 
in spite of their suppression by powers of $v^2$.  
After $|R(0)|^2$, the next most important matrix elements are 
$\langle {\cal O}^\psi_8(^3S_1) \rangle$,
$\langle {\cal O}^\psi_8(^1S_0) \rangle$, and
$\langle {\cal O}^\psi_8(^3P_0) \rangle$, all of which are 
suppressed by $v^4$.  They give the probabilities
for forming a $\psi$ from pointlike color-octet $c \bar c$ pairs
in spin-triplet S-wave, spin-singlet S-wave, and spin-triplet P-wave states, 
respectively.  These matrix elements are nonperturbative quantities that  
cannot be related in any simple way to the $c \bar c$ wavefunction.
We have no effective means of calculating them from first principles.
These matrix elements must therefore be treated as phenomenological 
parameters to be determined from experimental data. 

Now consider the fragmentation function $D_{g \to \psi}(z)$
that describes the formation of a  $\psi$ from a high-$p_T$ gluon.  
The color-singlet model term in the fragmentation function
has a short-distance factor of order $\alpha_s^3(m_c)$ and a 
long-distance factor that scales like $v^3$.  All other terms have 
long-distance factors that are suppressed
by powers of $v^2$, but there are some for which the short-distance 
factor is lower order in $\alpha_s$.  In particular, the term
that corresponds to the parton process $g \to c \bar c$ has a 
short-distance factor that is only of order $\alpha_s$.
The matrix element in this term is 
$\langle {\cal O}^\psi_8(^3S_1) \rangle$,
which gives the probability of forming a $\psi$ from a pointlike 
$c \bar c$ pair in a color-octet $^3S_1$ state. 
The fragmentation function at leading order in $\alpha_s$ is
\begin{equation}
D_{g \to \psi}(z) \;=\; 
{\pi \alpha_s(m_c) \over 24 m_c^3} \delta(1-z)
	\langle {\cal O}^\psi_8(^3S_1) \rangle.
\label{Dfrag}
\end{equation}
This term in the  fragmentation function is of order $\alpha_s v^7$, 
compared to $\alpha_s^3 v^3$ for the color-singlet term.   
It is not immediately 
obvious which of these two terms is more important,
but experience tells us that suppression by a power of $\alpha_s$ 
tends to be more effective than suppression by a power of $v^2$.
Thus it is not unreasonable for this color-octet term to dominate.

If the color-octet term is sufficiently large, it could provide an 
explanation for the CDF data on prompt charmonium production
\cite{Braaten-Fleming}.
The color-octet term in  the cross section at large $p_T$ is obtained by
inserting the fragmentation function (\ref{Dfrag}) into (\ref{sigfrag}).
The resulting prediction for the differential cross section as a 
function of $p_T$ is shown as a solid line in Figure 1. Its 
shape fits the data quite well.
The matrix element $\langle {\cal O}^\psi_8(^3S_1) \rangle$
can be adjusted so that the curve also agrees with the data in normalization. 
The value of the matrix element that is required is 
$\langle {\cal O}^\psi_8(^3S_1) \rangle$ = 0.014 GeV$^3$.
Now for this explanation of the data to be viable,
the matrix element must satisfy an important consistency check.
It must be small enough to be consistent with suppression by $v^4$
relative to the corresponding color-singlet factor $|R(0)|^2$.
Translating both quantities into the probability density for a point-like
$c \bar c$ pair to bind to form a $\psi$, we get 21/fm$^3$ for 
the color-singlet $^3S_1$ state and 1.8/fm$^3$ for 
the color-octet $^3S_1$ state.  The latter value is 
perfectly consistent with suppression by a factor of $v^4$.

We have shown that a color-octet term in the gluon fragmentation 
function provides a plausible explanation for the CDF data 
on prompt $\psi$ production.  To make this explanation convincing, 
we have to show that this color-octet production mechanism 
explains other aspects of charmonium production.  There are two 
particularly dramatic predictions that should be tested experimentally
in the near future. The first prediction is that at the Tevatron,
$\psi'$'s at large $p_T$ should be almost completely 
transversely polarized \cite{Cho-Wise}.
The $\psi'$ essentially inherits the transverse
polarization of the fragmenting gluon.  There is an effort 
underway at CDF to measure this polarization.  The second prediction 
is that prompt $\psi$'s should be observable in 
$Z^0$ decay at LEP \cite{C-K-Y}.
The color-singlet model predicts a rate for prompt $\psi$ production
that is too small to be observed.  Using the value of 
$\langle {\cal O}^\psi_8(^3S_1) \rangle$ obtained by fitting the CDF data,
we find that the color-octet mechanism increases the 
production rate by almost an order of magnitude.  The predicted branching 
fraction for $Z^0 \to \psi + X$ is $1.4 \times 10^{-4}$, 
which is large enough that it should be observable.  
Preliminary measurements of prompt $\psi$ production
from some of the LEP detectors seem to indicate that the 
branching fraction does indeed have the magnitude predicted by
the color-octet production mechanism.

\section{CONCLUSIONS}

If we understand charmonum production in simple high energy collisions, 
it should be possible to predict the production rate in heavy-ion collisions.
A new framework has been developed for describing charmonium production 
in those kinematic regions where it is dominated by parton collisions.
The cross section is factored into parton cross sections that 
can be computed using perturbation theory and  well-defined matrix elements
that give the probability for forming the bound state. 
The matrix elements with the greatest phenomenological importance 
include $|R(0)|^2$, 
$\langle {\cal O}_8(^3S_1) \rangle$,
$\langle {\cal O}_8(^1S_0) \rangle$, and
$\langle {\cal O}_8(^3P_0) \rangle$ for $\psi$ and $\psi'$
and $|R'(0)|^2$ and 
$\langle {\cal O}_8(^3S_1) \rangle$ for $\chi_{cJ}$.
It should be possible to determine these factors with reasonable
precision by carrying out a thorough analysis of all the available data on 
charmonium production, 
including fixed target data from $pN$, $\pi N$ and $e N$ collisions 
as well as data from the high energy $p \bar p$, $e^+ e^-$, 
and $e p$ colliders.    
The resulting matrix elements can then be used as inputs for 
predictions of charmonium production in heavy-ion collisions.

In conclusion, new data from high energy colliders has been 
driving progress in heavy quarkonium physics.  There are powerful 
theoretical tools that can be used to study heavy quarkonium.
The developments described above have exploited two of these tools:
perturbation theory 
and the relativistic expansion.  The multipole expansion
is another powerful tool that has important implications for 
heavy quarkonium production \cite{Kharzeev}.
By exploiting these tools, it should be possible to understand
heavy quarkonium in sufficient detail to allow it to be used 
as a probe in heavy-ion collisions.


\begin{thebibliography}{9}

\bibitem{B-F-Y}
E. Braaten, S. Fleming, and T.C. Yuan, OHSTPY-HEP-T-96-001 (hep-ph/9602374), 
to appear in Annual Reviews of Nuclear and Particle Science.

\bibitem{Schuler} 
G.A. Shuler, CERN-TH.7170/94 (hep-ph/9403387),
to appear in Physics Reports.

\bibitem{Sansoni}
A. Sansoni et al. (CDF Collaboration), 
FERMILAB-CONF-95/263-E, to appear in the 
Proceedings of the Sixth International Symposium on Heavy Flavor Physics.

\bibitem{B-B-L} 
G.T. Bodwin, E. Braaten, and G.P. Lepage, Phys. Rev. {\bf D51}, 1125 (1995).

\bibitem{Braaten-Fleming} 
E. Braaten and S. Fleming, Phys. Rev. Lett. {\bf 74}, 3327 (1995).

\bibitem{Cho-Wise} 
P. Cho and M. Wise, Phys. Lett. {\bf B346}, 129 (1995);
M. Beneke and I.Z. Rothstein, UM-TH-95-23 (hep-ph/9509375).

\bibitem{C-K-Y} 
K. Cheung, W.-Y. Keung, and T.C. Yuan, Phys. Rev. Lett. {\bf 76}, 877 (1996);
P. Cho, Phys. Lett. {\bf B368}, 171 (1996).

\bibitem{Kharzeev} 
D. Kharzeev, CERN-TH-95-342 (nucl-th/9601029).

\end{thebibliography}
\end{document}